\documentclass[twoside]{ilcws08}
\usepackage[latin1]{inputenc}
\usepackage[dvips]{graphicx,epsfig,color}
\usepackage{wrapfig,rotating}
\usepackage{amssymb,amsmath,array,units}

\begin{document}
\title{
TPC Readout Electronics with Time-to-Digital Converters}
\author{Alexander Kaukher
\thanks{This work is supported by the EUDET project.}
\vspace{.3cm}\\
 Universit\"at Rostock - Institut f\" ur Physik\\
 Universit\"atsplatz 3, 18051 - Germany
}

\maketitle

\begin{abstract}
Development of readout electronics for Time Projection Chamber for a 
Linear Collider is ongoing under stringent requirements on high channel 
density, lowest possible power consumption and small material budget.
In the studied TPC readout electronics time and charge of TPC signals 
are measured with the help of Time-to-Digit Converters. Optimization of 
performance of this electronics is considered and a methodology of signal 
simulation is presented.
\end{abstract}

\section{Introduction}

\begin{wrapfigure}{r}{0.5\columnwidth}
\centerline{\includegraphics[width=0.5\columnwidth]{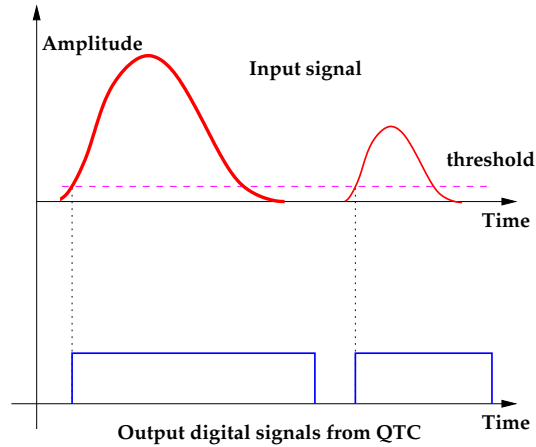}}
\caption{Working principle of the TPC readout electronics with TDCs.}
\label{Fig:TQ}
\end{wrapfigure}

The development of the TPC readout electronics with Time-to-Digit Converters(TDC) is carried out in the line of the Electronics Work-package of the LCTPC Collaboration\cite{Bellerive:2007}. Time of arrival and charge of signals from TPC pads are measured with the help of a time-to-digital converter. The charge is measured indirectly, with help of a charge-to-time converter(QTC) - the charge is encoded into the pulse width on the output of the QTC.\\

For smaller input signals, the output pulse width is shorter, Figure \ref{Fig:TQ}. In order to accommodate for large dynamic range of signals from TPC, it is convenient to use, for example, log-like charge-to-time conversion characteristics.

Current implementation\cite{eudetmemo08} of the readout electronics uses ASDQ chips. 
This circuit offers simple timing method with threshold discriminator. The method gives good results in 
applications with steep signal edges, like in proportional counters. For a GEM TPC detector, where signals are induced on pads, timing performance need to be optimized.\\

The subject of this work is the simulation of the signals induced on pads of a GEM TPC detector.

\section{Simulation of signals from a GEM TPC detector}
Derivation of signals induced on the pads of a GEM TPC detector can be done by using the Ramo's theorem\cite{Ramo:1939vr}:
\begin{equation}
{I = q~{\vec{\bf E}_w} \cdot {\vec{\bf v}}}
\end{equation}

\begin{wrapfigure}{r}{0.5\columnwidth}
\centerline{\includegraphics[width=0.5\columnwidth]{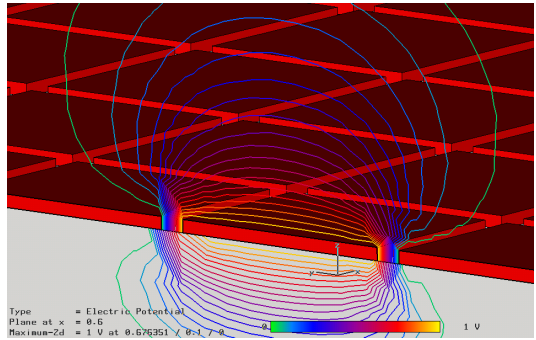}}
\caption{Weighting field is visualized in form of potential lines on a cutting plane. On this figure z-axis is pointing up.}
\label{Fig:WField}
\end{wrapfigure}

Current $I$ is induced in a pad by charge $q$ moving with velocity $\vec{ \bf v}$. 
Here charge is considered to be point-like. ${\vec{\bf E}_w}$ is the weighting field. The weighting field is 
defined by unit potential on the pad under consideration in the absence of the charge, while all other 
surrounding electrodes are at zero potential. Effectively, the weighting field shows coupling of a charge to the pad. 

Convenient way to calculate the weighting field in a multi-electrode configuration is to use a Finite Element Method.
In this work {\em CST Studio Suite\texttrademark}\cite{CST} was used.
An example of the weighting field is shown on Figure \ref{Fig:WField}. 
The weighting field visualized in the form of potential lines on a vertical plane cutting a pad at the middle. 
In this example, the pad has squared shape. Note how the potential lines gets denser in vicinity of the pad.

In a real GEM TPC, signals typically induced by a cluster of charges with finite size, therefore one would 
need to combine results of individual signal inductions. Position of individual charges in a cloud can be found 
with the help of a TPC simulation program, for example Marlin-TPC\cite{Abernathy:2007gk}.
Hereinafter only point-like charges will be considered.
\begin{wrapfigure}{r}{0.5\columnwidth}
\centerline{\includegraphics[width=0.5\columnwidth]{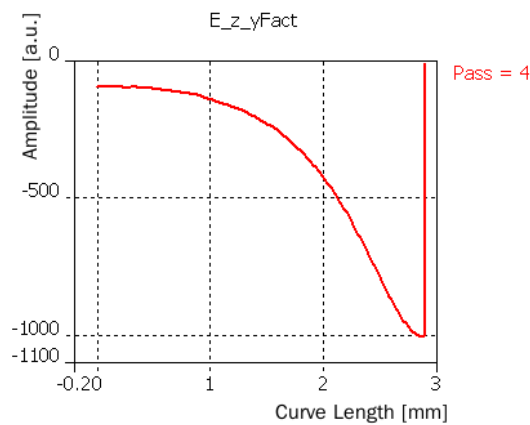}}
\caption{A signal from a point-like charge on a TPC pad. The induction gap is \unit[3]{mm}.}
\label{Fig:Signal}
\vspace{-0.9cm}
\end{wrapfigure}

A charge moves under the influence of the electric field applied between the GEM surface and the pads.
Drift velocity (and therefore drift time) of the charge depends on the magnitude of the electric field. 
The drift time define the period of signal development and it is typically of the order of \unit[50]{ns}.\\

Assume the charge drifts without diffusion, strictly along the z-axis, Figure \ref{Fig:WField}. In this case, the shape 
of the induced signal is given by the z-component of the weighting field, Figure \ref{Fig:Signal}.
Note the signal peaks as the charge approaches the pad. Compared to proportional counters, signal from 
a GEM detector has an order of magnitude longer front edge. This implies that in the current implementation 
of the readout electronics may not necessary be optimal for this application, so that the timing 
performance might be insufficient. 

One of the possibilities to improve timing performance would be to reduce the size of the induction gap, reducing the drift 
time of the charges. However this might negatively effect performance of the coordinate reconstruction in the x-y plane.
Another possibility is to study other methods of timing, for example, peak detecting circuit.\\

Integral of the induced current gives the total accumulated charge on the pad. For different y-positions of the charge 
one can obtain different waveforms and therefore - set of signal charges induced on the pad. This gives direct input 
to the calculation of the Pad Response Function of the pad.

\section{Conclusion}
Further development of the TPC readout electronics will require optimization of its performance.
For this purpose, GEM TPC signal simulation is being prepared and will have to be combined with a TPC simulation package.

It is planned to study GEM signals from an existing small GEM TPC detector. This will help to verify the 
simulation of GEM signals. This can be performed with a small GEM chamber with short drift distance.

As a methods of timing a peak detecting circuit can be considered.
% Furthermore, the double pulse capability of the charge-to-time converter will need to be optimized.

\begin{footnotesize}

\end{footnotesize}
\end{document}